\documentclass[acmsmall]{acmart}

\copyrightyear{2024}
\acmYear{2024}
\setcopyright{rightsretained}
\acmConference[Scheme 2024]{Proceedings of the 2024 Workshop on Scheme and Functional Programming}{September 7, 2024}{Milan, Italy} \acmBooktitle{Proceedings of the 2024 Workshop on Scheme and Functional Programming, September 7, 2024, Milan, Italy} 

\usepackage{balance}
\usepackage[utf8]{inputenc} 
\usepackage{graphicx} 

\usepackage{float}
\newcommand\klammeraffe{\makeatletter@\makeatother }

\title[Beyond SICP---A Notional Machine for Scheme]{Beyond SICP---Design and Implementation of a Notional Machine for Scheme}

\author{Kyriel Mortel Abad}
\authornote{All authors contributed equally to this research.}
\email{kyriel@u.nus.edu}
\affiliation{%
  \institution{National University of Singapore}
  \country{Singapore}
   \city{Singapore} 
}

\author{Martin Henz}
\authornotemark[1]
\email{henz@comp.nus.edu.sg}
\affiliation{%
  \institution{National University of Singapore}
  \country{Singapore}
   \city{Singapore} 
}

\date{\today}


\makeatletter
\gdef\@copyrightpermission{
   \begin{minipage}{0.3\columnwidth}
     \href{https://creativecommons.org/licenses/by-sa/4.0/}{\includegraphics[width=0.90\textwidth]{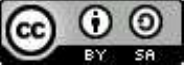}}
   \end{minipage}\hfill
   \begin{minipage}{0.7\columnwidth}
     \href{https://creativecommons.org/licenses/by-sa/4.0/}{This work is licensed under a Creative Commons Attribution-ShareAlike International 4.0 License.}
   \end{minipage}
   \vspace{5pt}
}
\makeatother

\begin{document}

\begin{abstract}
Computer science education has been at the heart of Scheme from the beginning. The language was designed in the 1970s concurrently with the MIT course 6.001 and the
textbook ``Structure and Interpretation of Computer Programs'' (SICP). To explain the scope of variables at run time in the presence of higher-order procedures, SICP introduces a mental model called the \emph{environment model}, along with a pictorial representation of environments and data structures. Recently, the concept of \emph{notional machines} has emerged in computer science education: a predictive set of abstractions that describe the structure and behavior of a computational device. Proponents of notional machines argue that learners benefit when complex dynamic concepts such as the computational structure of Scheme are accompanied by concise notional machines. 
In this paper, we start with a sublanguage of Scheme sufficient for all programs in SICP that we call \emph{SICP Scheme}. We extend the environment model to a full notional machine for SICP Scheme that is simple enough to serve as the central mental model in a CS1 course, and demonstrate the machine with computer-generated visualizations. Moving beyond SICP Scheme, we show how the notional machine can be further extended to explain Scheme's \texttt{call/cc} and thus make this powerful concept accessible to beginners through a coherent mental model. The presented notional machine serves as the core of a web-based implementation of Scheme that is under development at the National University of Singapore.
\end{abstract}


\maketitle

\section{Introduction}
\label{sec:background}

Mental models are cognitive representations that allow individuals to understand, predict, and reason about a system or phenomenon~\cite{mental_models}. Pedagogical research in various scientific disciplines, including physics, chemistry, and biology, has highlighted the importance of mental models in the learning process, as they help students organize knowledge, generate explanations, and solve problems~\cite{mental_models_science}. In computing, the development of proper mental models enables a learner to advance from superficial familiarity with a subject to the deep understanding required for innovation. Introductory computer science courses (CS1) provide the opportunity to emphasize their importance in general by giving adequate mental models for the behavior of programs as examples. 

A particular family of mental models called \emph{notional machines}~\cite{notional} have recently gained prominence in computer science. As Guzdial states~\cite{notional_guzdial}, ``[A] notional machine (NM) [is] a set of abstractions that define the structure and behavior of a computational device. A notional machine includes a grammar and a vocabulary, and is specific to a programming paradigm. It's consistent and predictive---given a notional machine and a program to run on that machine, we should be able to define the result.'' A notional machine can be seen as an operational semantics that is tailor-made for learners of a programming paradigm.

Pioneering educators such as Donald Knuth and Edsger W. Dijkstra recognized the importance of providing students with a clear and consistent mental model of program execution. Knuth, in his seminal work \emph{The Art of Computer Programming}~\cite{knuth97}, presented a detailed and systematic approach to programming, using the MIX assembly language. Dijkstra, in his influential book \emph{A Discipline of Programming}~\cite{books/ph/Dijkstra76}, advocated for a structured approach to programming, such that the programmer is able to construct a \emph{weakest precondition} for any program fragment and postcondition. Both approaches can be viewed as notional machines because they let the learner systematically predict the result of executing any given program in the respective framework.

In the influential textbook \emph{The Structure and Interpretation of Computer Programs} (SICP)~\cite{sicp}, Abelson and Sussman introduce two mental models for understanding the behavior of programs in the Scheme programming language: the substitution model and the environment model. The substitution model of Chapters 1 and 2 provides an intuitive understanding of how iterative and recursive processes in purely functional languages work by systematically replacing formal parameters with their actual values. To go beyond purely functional programming, Chapter 3 introduces the environment model, which explains how lexical scoping works at run time by maintaining a chain of environments that map variable names to their values. Although neither the substitution model nor the environment model is fully developed into a notional machine in SICP, they provide a foundation for understanding the behavior of programs in a language with first-class functions and lexical scoping. The final Chapter 5 provides two notional machines for program execution, an interpreter and a compiler, both based on a register machine. Unfortunately, the low-level nature of the register machine prevents the use of these notional machines as a mental model to explain the execution of actual programs in the central chapters, a problem that Knuth solved with the consistent use of MIX.

In recent years, introductory computer science courses, such as 6.0001 at MIT and CS 61A at Berkeley, have adopted the programming language Python as the language of choice, replacing Scheme. These courses often incorporate elements from SICP such as the environment model to help students understand the behavior of programs. Tools like Python Tutor~\cite{pythontutor}, which provides a visual representation of program execution, have been employed to aid students in understanding how Python programs work. While these tools offer valuable insights into the state of variables and the flow of control, they do not provide a notional machine that captures Python's first-class functions and expression evaluation. For example, the role of a ``run-time stack'' as propagated by Python tutor is problematic in the context of higher-order functions with lexical scoping, as pointed out by Clements and Krishnamurthi~\cite{stacks_dont_stack}.

The absence of a comprehensive notional machine for high-level programming languages can hinder students' ability to mentally trace the execution of their programs, leading to difficulties in understanding, debugging, and predicting the outcome of their code. As Guzdial emphasizes, ``given a notional machine and a program to run on that machine, we should be able to define the result.'' More specifically, given a machine state, students should be able to predict the next step in the program's execution. Without a notional machine that captures the essential characteristics of modern languages, students are left to rely on vague intuitions instead of a solid semantic foundation for the central features of these languages. 

Most of these features are of course inherited from Scheme: lexical scoping, first-class procedures, variable assignment, mutable data structures, dynamic typing, and memory safety. It seems appropriate to us to try to develop a notional machine for Scheme that would then allow for an adaptation to recent variants such as Python. Our notional machine for Scheme should be simple enough to serve as a structured and unified framework for understanding program execution. By equipping students with a clear and consistent mental model of how programs are executed at each step, a notional machine for Scheme would enable students to accurately predict the behavior of their programs, facilitating a deeper understanding of the language semantics and promoting more effective learning and teaching of fundamental programming concepts. Our pedagogical goal is a notional machine for all programs in SICP, so initially, we focus on the SICP sublanguage of Scheme, which we call SICP Scheme. 

\section{Background and Overview}
\label{sec:requirements}

The lambda calculus~\cite{barendregt} is a reasonable starting point to capture Scheme's first-class procedures and lexical scoping.  From the point of view of programming language research, most notional machines are small step operational semantics designed for teaching. The classical small-step semantics for the applicative-order lambda calculus is Landin's SECD machine~\cite{Landin1964TheME}; Danvy~\cite{DBLP:conf/ifl/Danvy04} systematically analyses a family of SECD variants. Based on John Reynolds's interpreters~\cite{10.1145/800194.805852}, Felleisen and Friedman developed the CEK machine~\cite{redex} that simplifies the SECD machine by combining its environment E with a context component C and a continuation component K, which replace SECD's stack S, control C and dump D. In these frameworks, it is common to use de Bruijn indices~\cite{de-Bruijn:IM72} instead of named variables so that environments can be represented by simple lists.

Different simplifications of the SECD machine appeared more recently in the context of correctness proofs for small-step semantics for the applicative-order lambda calculus. The ``Modern SECD Machine'' described by Xavier Leroy~\cite{modern} combines the dump D component with a global operand stack S. Another approach proposed by Kunze, Smolka, and Forster~\cite{kunze_aplas2018} combines the dump D component and the control C component in their ``closure machine''. The use of de Bruijn indices requires a compilation step prior to program execution, which provides a significant stumbling block for learners. To our knowledge, none of these approaches has been used as mental models in first-year programming.

The next section develops a notional machine for SICP Scheme loosely based on Kunze, Smolka, and Forster's closure machine. To achieve a purely interpretative approach, we refrain from using de Bruijn indices. We shall call the result the \emph{CSE machine}.
Section~\ref{sec:call-cc} demonstrates how Scheme's \texttt{call/cc} construct fits naturally into the CSE machine.
To our knowledge, the Computer Science Departments at Uppsala University and the National University of Singapore are the only major universities worldwide that are teaching their CS1 course using SICP. Their courses CS1101S and PKD, respectively, follow the JavaScript edition~\cite{sicpjs} of SICP. CS1101S covers all its chapters, including the metacircular evaluator of Chapter 4 and the register machine of Chapter 5. 
The needs of educators that follow SICP JS led to the development of Source Academy~\cite{sa}, an open-source programming and learning environment sustained by a community of students and educators~\cite{anderson_sigcse2023}. Section~\ref{sec:implementation} explains how our implementation fits into the Source Academy implementation. Source Academy comes with a visualizer for SICP-style environments~\cite{env-visu-2023} that we have adapted and extended for the purpose of visualizing Scheme programs. Section~\ref{sec:visualization} shows how the CSE machine is visually represented to the learner. We conclude with a vision for an extended SICP pedagogy in which the CSE machine serves as the central notional machine for a Scheme-based CS1 course.

\section{A Notional Machine for SICP Scheme}
\label{sec:notional}

In this and the next section, we define a one-step reduction relation $\rightarrow$ or $\downarrow$ as a partial function from a state to the next state, and $\rightarrow^*$ or $\downarrow^*$ as its reflexive transitive closure.

We propose a notional machine for SICP Scheme---the \emph{CSE machine}---that represents the evaluation of a SICP Scheme program as a sequence of states,
where at each step of the evaluation, the next state is fully determined by the current state. The state is a triple $(C, S, E)$, where $C$ is called the \emph{control}, $S$ is called the \emph{stash}, and $E$ is the environment from Chapter 3 of SICP.

In more detail, the stash\footnote{Traditionally, this component is called ``operand stack'' or simply ``stack''. Considering that $C$ is also a stack, we take the liberty to introduce a new terminology that seems to work well in our first-year course.} $S$ is a stack (list) of run-time values such as numbers, symbols, and pairs. In the rules below, we use $\epsilon$ as the empty list and $:$ as an infix pair constructor, so $3 : 4 : \epsilon$ is a list that contains the two values 3 and 4. The environment $E$ maps variables to values and an empty environment is represented ambiguously by $\epsilon$. We denote the lookup of a program variable $x$ in an environment $E$ by $E.x$, which results in the value to which the variable $x$ is bound in $E$. We denote the extension of an environment $E$ by a binding of $x$ to a value $v$ by $E[x = v]$. 

We express the destructive update of environment $E$ by a binding of $x$ to a new value $v$ as $E[x \leftarrow v]$. This destructive update introduces a major conceptual problem in the semantic framework as presented in this section. Environments get copied in some rules, and destructive updates need to apply to all previously obtained copies! Fortunately, the graphical notation introduced in Section~\ref{sec:visualization} accurately reflects structure sharing in the presence of destructive updates. A proper handling of destructive update would necessitate a representation of the state as a graph that formalizes the visualization presented to the learners. In this and the next section, we ignore this issue for the sake of readability of the rules.

The control $C$ is a stack (list) of components, which are either Scheme expressions or \emph{instructions}. Instructions are control components that the CSE machine generates during program execution. We use all-caps for constructors, for example the nullary constructor $\textsc{pop}$. As with stashes, we use the infix pair constructor $:$ and $\epsilon$ for the empty control.

Our CSE machine evaluates a given Scheme program $p$ by repeatedly applying the rules in this and the following section to the initial state $(p:\epsilon, \epsilon, E_0)$ until a final state $(\epsilon, v:\epsilon, E)$ is reached. The purpose of the rules in this paper is to concisely capture the intended small-step semantics for the benefit of the programming language researcher and educator; they are not presented to the learners. In our pedagogical approach, the learners are absorbing the workings of the CSE machine by observing progression of states using the graphical notation presented in Section~\ref{sec:visualization}.

The initial environment $E_0$ contains bindings for primitive operations and the single value $v$ on the final stack is the result of evaluation of $p$. At any time, the first component on the control uniquely determines which rule applies.

\begin{gather*}
(v:C, S, E) \quad \rightarrow \quad (C, v:S, E) \\
\textrm{Evaluation of a primitive value} \ v \ \textrm{such as a number, symbol, boolean, or string} \\
(x:C, S, E) \quad \rightarrow \quad (C, E.x:S, E) \\
\textrm{Lookup of variable}\ x\ \textrm{in environment} \ \mathit{E} \\
\end{gather*}

All Scheme expressions are decomposed into lists of subexpressions and instructions. Five different instructions facilitate the CSE machine operations, namely \textsc{asgn}, \textsc{call}, \textsc{env}, \textsc{branch} and~\textsc{pop}. 

A Scheme procedure application is decomposed into a sequence of its subexpressions denoted by $V_0 \ldots V_n$, followed by a \textsc{call} instruction.

\begin{gather*}
(\texttt{(} V_0\ V_1 \ldots  V_n \texttt{)}:C, S, E) \quad \rightarrow \quad (V_0: V_1:\ldots:V_n:\textsc{call}\ n:C, S, E) \\
\textrm{Decomposition of} \ n\textrm{-ary Scheme procedure call} \\
\end{gather*}

Execution of the \textsc{call} instruction carries out the application. \textsc{call} fulfils the same role as the special object $\textit{ap}$ in~\cite{Landin1964TheME}, but takes the procedure's arity as a numeric parameter whereas Landin only considers unary functions in the SECD machine. This rule enforces left-to-right evaluation of procedure arguments, whereas Scheme~\cite{r7rs} leaves the evaluation order unspecified.

\begin{gather*}
(\textsc{call}\ n:C, v_n:\ldots:v_1:f:S, E) \quad \rightarrow \quad (C, f(v_1,\ldots,v_n):S, E) \\
\textrm{Call of simple procedure or operator}\ f \\
\end{gather*}

The following machine run illustrates how  execution proceeds using the expression \texttt{(* 2 3)} as an example, assuming that the multiplication operator $\textit{times}$ is bound to the variable $\texttt{*}$ in $E_0$. In the following sections, we assume that the initial environment $E_0$ contains appropriate bindings for arithmetic operations such as $\texttt{*}$, $\texttt{/}$, $\texttt{+}$ and $\texttt{-}$, as well as logical operations such as $\texttt{<}$, $\texttt{>}$ or $\texttt{=}$.

\begin{gather*}
(\texttt{(* 2 3)}:\epsilon, \epsilon, E_0) \\
\downarrow \\
(\texttt{*}:\texttt{2}:\texttt{3}:\textsc{call}\ 2:\epsilon, \epsilon, E_0) \\
\downarrow \\
(\texttt{2}:\texttt{3}:\textsc{call}\ 2:\epsilon, \textit{times}:\epsilon, E_0) \\
\downarrow \\
(\texttt{3}:\textsc{call}\ 2:\epsilon, 2:\textit{times}:\epsilon, E_0) \\ 
\downarrow \\
(\textsc{call}\ 2:\epsilon, 3:2:\textit{times}:\epsilon, E_0) \\ 
\downarrow \\
(\epsilon, 6:\epsilon, E_0) \\ 
\end{gather*}

The \textsc{call} instruction can also be applied to user-defined procedures. In our CSE machine, we represent a procedure closure using a constructor $\textsc{clo}$ as in $\textsc{clo}\ (x_1 \ldots x_n)\ B\ E$, where $(x_1 \dots x_n)$ are the parameters of the procedure, $B$ is the body of the procedure and $E$ is the environment to be used in applications.

\noindent
\begin{minipage}{\textwidth}
\begin{gather*}
(\texttt{(lambda\ (}x_1  \ldots  x_n\texttt{)} \ B \texttt{)}:C, S, E) \quad \rightarrow \quad (C, \textsc{clo} \ (x_1,\ldots, x_n) \ B \ E:S, E) \\
\textrm{Construction of a procedure closure from a lambda expression}
\end{gather*}
The two occurrences of $E$ in the new state on the right are not meant as independent copies, but as shared nodes in a graph, as will become visible in Section~\ref{sec:visualization}.
\end{minipage}

\begin{gather*}
(\textsc{call}\ n:C, v_n:\ldots:v_1:\textsc{clo} \ (x_1,\ldots,x_n) \ B \ E':S, E) 
\rightarrow 
(B:\textsc{env}\ E:C, S, E'[x_1 = v_1]\ldots[x_n = v_n]) \\
\textrm{Call of procedure closure with arguments} \ v_1  \ldots  v_n 
\end{gather*}
The instruction \textsc{env} in the new control restores the original environment $E$ after execution of the procedure body $B$ in the procedure closure's environment $E'$ extended by bindings of the arguments to the parameters.

\begin{gather*}
(\textsc{env}\ E':C, S, E) \quad \rightarrow \quad (C, S, E') \\
\textrm{Environment restoration}
\end{gather*}

\begin{figure}
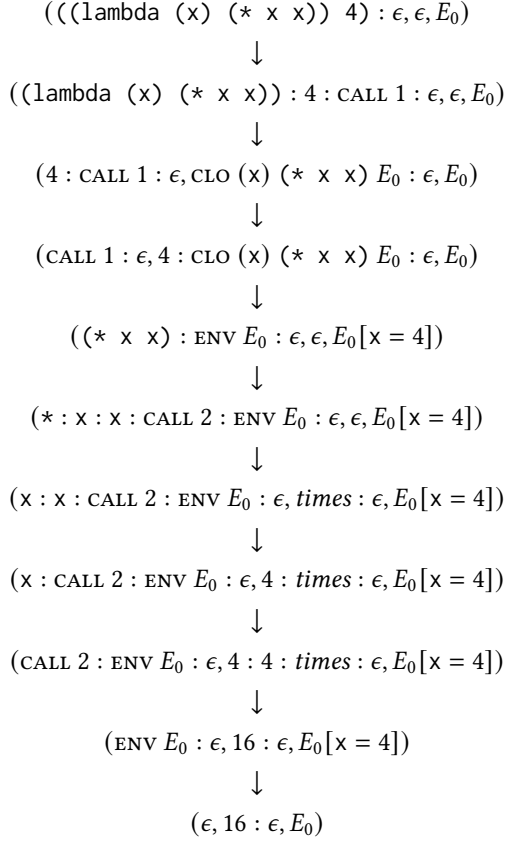

\begin{gather*}
(\texttt{((lambda (x) (* x x)) 4)}:\epsilon, \epsilon, E_0) \\
\downarrow\\
(\texttt{(lambda (x) (* x x))}:4:\textsc{call}\ 1 : \epsilon, \epsilon, E_0) \\
\downarrow\\
(4:\textsc{call}\ 1:\epsilon, \textsc{clo} \ (\texttt{x}) \ \texttt{(* x x)} \ E_0:\epsilon, E_0) \\
\downarrow\\
(\textsc{call}\ 1:\epsilon, 4:\textsc{clo} \ (\texttt{x}) \ \texttt{(* x x)} \ E_0:\epsilon, E_0) \\
\downarrow\\
(\texttt{(* x x)}:\textsc{env} \ E_0:\epsilon, \epsilon, E_0[\texttt{x} = 4]) \\
\downarrow\\
(\texttt{*}:\texttt{x}:\texttt{x}:\textsc{call}\ 2:\textsc{env} \ E_0:\epsilon, \epsilon, E_0[\texttt{x} = 4]) \\
\downarrow\\
(\texttt{x}:\texttt{x}:\textsc{call}\ 2:\textsc{env} \ E_0:\epsilon, \textit{times}:\epsilon, E_0[\texttt{x} = 4]) \\
\downarrow\\
(\texttt{x}:\textsc{call}\ 2:\textsc{env} \ E_0:\epsilon, 4:\textit{times}:\epsilon, E_0[\texttt{x} = 4]) \\
\downarrow\\
(\textsc{call}\ 2:\textsc{env} \ E_0:\epsilon, 4:4:\textit{times}:\epsilon, E_0[\texttt{x} = 4]) \\
\downarrow\\
(\textsc{env} \ E_0:\epsilon, 16:\epsilon, E_0[\texttt{x} = 4]) \\
\downarrow\\
(\epsilon, 16:\epsilon, E_0)
\end{gather*}
\caption{\label{fig:square}Creating and applying procedure closures in the CSE machine}
\end{figure}

Figure~\ref{fig:square} demonstrates the usage of procedure closures in the CSE machine by applying a square procedure. Note that the $\textsc{env}\ E_0$ instruction is not necessary in this example because $E_0$ is not used after restoration.

Variable declarations and assignments utilise the \textsc{asgn} instruction in their decomposition.

\begin{gather*}
(\texttt{(define} \ x \ V \texttt{)}:C, S, E) \quad \rightarrow \quad (V:\textsc{asgn} \ x:C, S, E) \\
\textrm{Decomposition of variable declaration} \\
(\texttt{(set!} \ x \ V \texttt{)}:C, S, E) \quad \rightarrow \quad (V:\textsc{asgn} \ x:C, S, E) \\
\textrm{Decomposition of variable assignment} \\
\end{gather*}

Note that in this implementation of Scheme in the CSE machine, \texttt{define} and \texttt{set!} are value-producing expressions.

The \textsc{asgn} instruction destructively updates a binding $x$ with a new value $v$ found at the top of the stack within the environment $E$.

\begin{gather*}
(\textsc{asgn} \ x:C, v:S, E) \quad \rightarrow \quad (C, v:S, E[x \leftarrow v]) \\ 
\textrm{Assignment of value} \ \mathit{v} \ \textrm{to variable x in environment} \ \mathit{E} \\
\end{gather*}
The graph structure of the state that is presented visually to the learner ensures that this destructive change affects all references to $E$ obtained by previous environment extensions and the construction of $\textsc{env}$ instructions and closures.

The example in Figure~\ref{fig:defsquare} defines the 
variable $\texttt{square}$ and binds it to a procedure closure in the resulting environment.

\newcommand{\verteq}{\mathrel{\rotatebox{90}{$=$}}}

\begin{figure}[t]
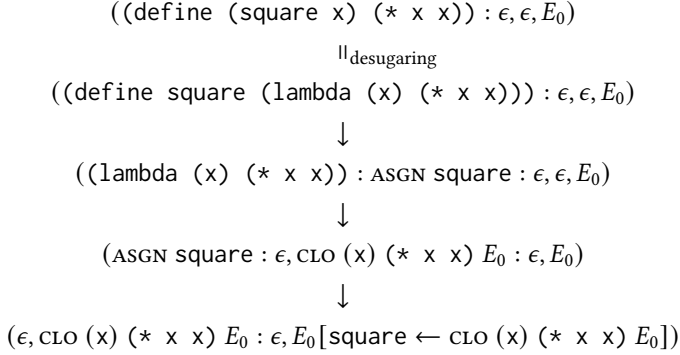

\begin{gather*}
(\texttt{(define (square x) (* x x))}:\epsilon, \epsilon, E_0) \\
\hspace{12.25mm}\verteq_{\textrm{desugaring}}\\
(\texttt{(define square (lambda (x) (* x x)))}:\epsilon, \epsilon, E_0) \\
\downarrow\\
(\texttt{(lambda (x) (* x x))}:\textsc{asgn} \ \texttt{square}:\epsilon, \epsilon, E_0) \\
\downarrow\\
(\textsc{asgn} \ \texttt{square}:\epsilon, \textsc{clo} \ (\texttt{x}) \ \texttt{(* x x)} \ E_0:\epsilon, E_0) \\
\downarrow\\
(\epsilon, \textsc{clo} \ (\texttt{x}) \ \texttt{(* x x)} \ E_0:\epsilon, E_0[\texttt{square} \leftarrow \textsc{clo} \ (\texttt{x}) \ \texttt{(* x x)} \ E_0]) 
\end{gather*}
\caption{\label{fig:defsquare} Defining a \texttt{square} procedure in the CSE machine}
\end{figure}

Conditional expressions in Scheme require an instruction that can handle the concept of conditional program flow, which is fulfilled with the \textsc{branch} instruction.

\begin{gather*}
(\texttt{(if} \ V \ A \ B \texttt{)}:C, S, E) \quad \rightarrow \quad (V:\textsc{branch}\ A \ B:C, S, E) \\
\textrm{Decomposition of conditional expression} \\
\end{gather*}

The \textsc{branch} instruction contains two Scheme expressions $A$ and $B$, of which only one will be evaluated based on the truthiness of the topmost value of the stack at the current state of evaluation.

\begin{gather*}
(\textsc{branch}\ A \ B:C, v:S, E) \quad \rightarrow \quad (A:C, S, E) \\
\textrm{Evaluation of} \ \textsc{branch} \ \textrm{instruction if $v$ is truthy} \\
(\textsc{branch}\ A \ B:C, v:S, E) \quad \rightarrow \quad (B:C, S, E) \\
\textrm{Evaluation of} \ \textsc{branch} \ \textrm{instruction if $v$ is falsy} \\
\end{gather*}

\begin{figure}[t]
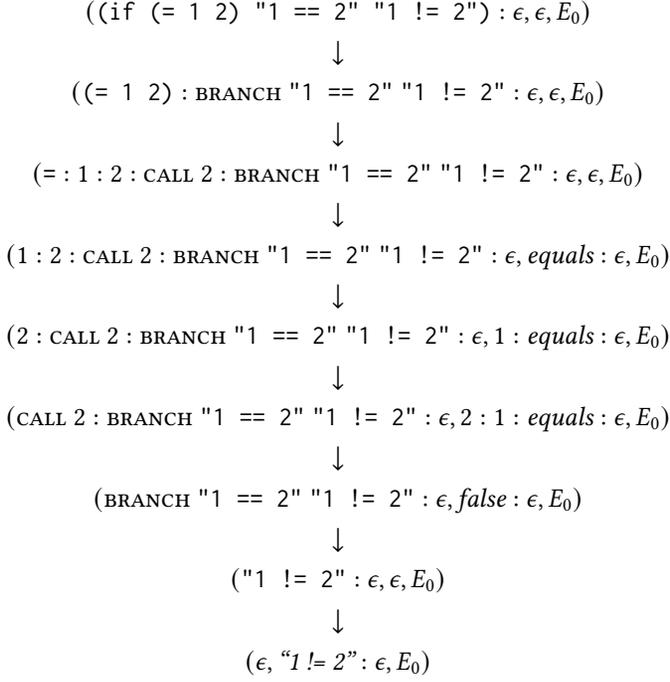

\begin{gather*}
(\texttt{(if (= 1 2) "1 == 2" "1 != 2")}:\epsilon, \epsilon, E_0) \\
\downarrow\\
(\texttt{(= 1 2)}: \textsc{branch} \ \texttt{"1 == 2"}\ \texttt{"1 != 2"}:\epsilon, \epsilon, E_0) \\
\downarrow\\
(\texttt{=}:1:2:\textsc{call}\ 2: \textsc{branch} \ \texttt{"1 == 2"}\  \texttt{"1 != 2"}:\epsilon, \epsilon, E_0) \\
\downarrow\\
(1:2:\textsc{call}\ 2: \textsc{branch} \ \texttt{"1 == 2"}\ \texttt{"1 != 2"}:\epsilon, \textit{equals}:\epsilon, E_0) \\
\downarrow\\
(2:\textsc{call}\ 2: \textsc{branch} \ \texttt{"1 == 2"}\ \texttt{"1 != 2"}:\epsilon, 1:\textit{equals}:\epsilon, E_0) \\
\downarrow\\
(\textsc{call}\ 2: \textsc{branch} \ \texttt{"1 == 2"}\ \texttt{"1 != 2"}:\epsilon, 2:1:\textit{equals}:\epsilon, E_0) \\
\downarrow\\
(\textsc{branch} \ \texttt{"1 == 2"}\ \texttt{"1 != 2"}:\epsilon, \textit{false}:\epsilon, E_0) \\
\downarrow\\
(\texttt{"1 != 2"}:\epsilon, \epsilon, E_0) \\
\downarrow\\
(\epsilon, \textit{``1 != 2''}:\epsilon, E_0) 
\end{gather*}
\caption{\label{fig:branch} Branching the CSE machine}
\end{figure}

Figure~\ref{fig:branch} demonstrates the behaviour of 
comparison operators and the \textsc{branch} instruction.

The \textsc{pop} instruction is used to remove values from the top of the stash that are no longer required for further computation, such as values produced during the evaluation of a sequence of expressions.

\begin{gather*}
(V_1  \ldots  V_n:C, S, E) \quad \rightarrow \quad (V_1:\textsc{pop}:\ldots:V_n:C, S, E) \\
\textrm{Decomposition of a sequence of expressions} \\
(\texttt{(begin} \ V_1  \ldots  V_n \texttt{)}:C, S, E) \quad \rightarrow \quad (V_1:\textsc{pop}:\ldots:V_n:C, S, E) \\
\textrm{Decomposition of a sequence of expressions with the} \ \texttt{begin} \ \textrm{syntax} \\
(\textsc{pop}:C, v:S, E) \quad \rightarrow \quad (C, S, E) \\
\textrm{Removal of an unused value} \\
\end{gather*}

\begin{figure}[tbp]
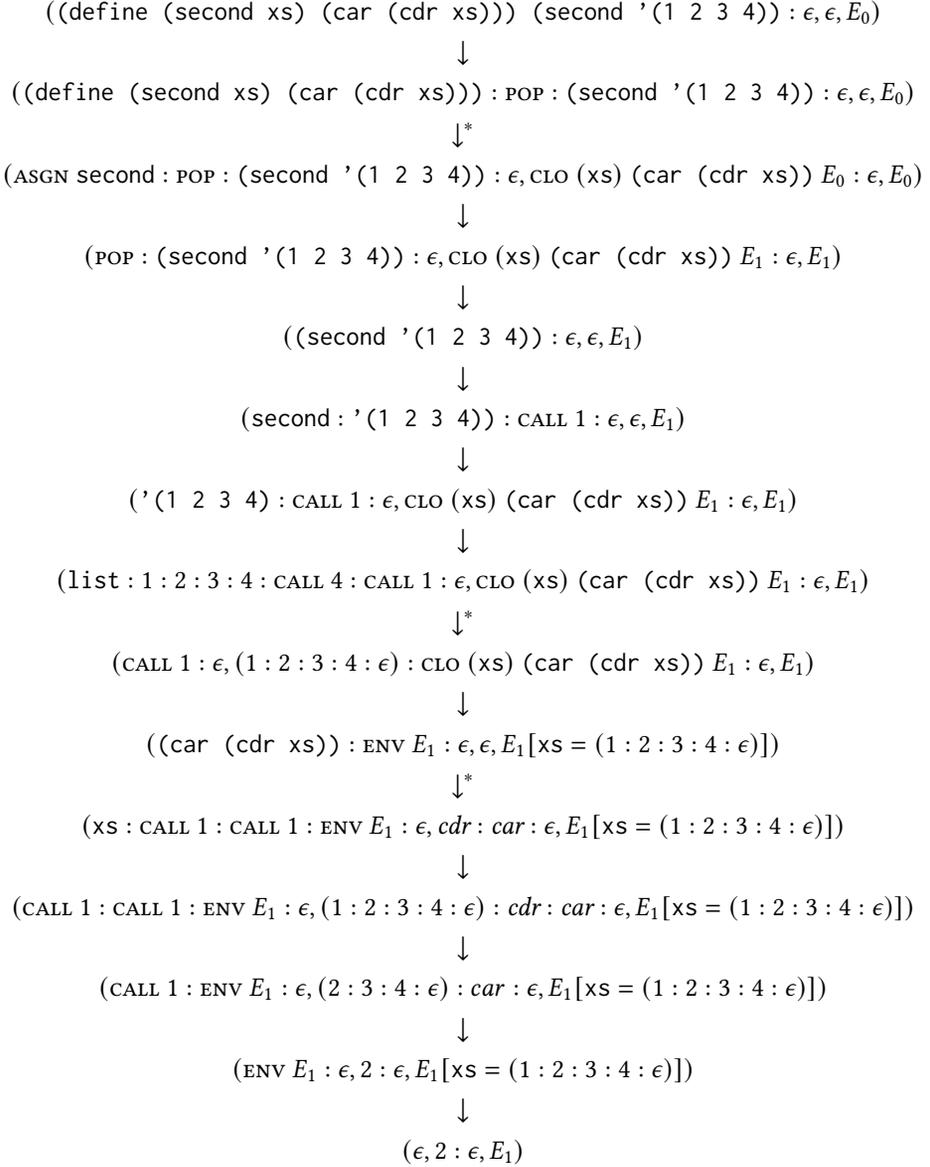

\begin{gather*}
(\texttt{(define (second xs) (car (cdr xs))) (second '(1 2 3 4))}:\epsilon, \epsilon, E_0)\\
\downarrow\\
(\texttt{(define (second xs) (car (cdr xs)))}:\textsc{pop}:\texttt{(second '(1 2 3 4))}:\epsilon, \epsilon, E_0)\\
\downarrow^*\\
(\textsc{asgn}\ \texttt{second}:\textsc{pop}:\texttt{(second '(1 2 3 4))}:\epsilon, \textsc{clo} \ (\texttt{xs}) \ \texttt{(car (cdr xs))} \ E_0:\epsilon, E_0)\\
\downarrow\\
(\textsc{pop}:\texttt{(second '(1 2 3 4))}:\epsilon, \textsc{clo} \ (\texttt{xs}) \ \texttt{(car (cdr xs))} \ E_1:\epsilon, E_1)\\
\downarrow\\
(\texttt{(second '(1 2 3 4))}:\epsilon, \epsilon, E_1)\\
\downarrow\\
(\texttt{second}:\texttt{'(1 2 3 4))}:\textsc{call}\ 1:\epsilon, \epsilon, E_1)\\
\downarrow\\
(\texttt{'(1 2 3 4)}:\textsc{call}\ 1:\epsilon, \textsc{clo} \ (\texttt{xs}) \ \texttt{(car (cdr xs))} \ E_1:\epsilon, E_1)\\
\downarrow\\
(\texttt{list}:1:2:3:4:\textsc{call}\ 4:\textsc{call}\ 1:\epsilon, \textsc{clo} \ (\texttt{xs}) \ \texttt{(car (cdr xs))} \ E_1:\epsilon, E_1)\\
\downarrow^*\\
(\textsc{call}\ 1:\epsilon, (1:2:3:4:\epsilon):\textsc{clo} \ (\texttt{xs}) \ \texttt{(car (cdr xs))} \ E_1:\epsilon, E_1)\\
\downarrow\\
(\texttt{(car (cdr xs))}:\textsc{env} \ E_1:\epsilon, \epsilon, E_1[\texttt{xs} = (1:2:3:4:\epsilon)])\\
\downarrow^*\\
(\texttt{xs}:\textsc{call}\ 1:\textsc{call}\ 1:\textsc{env} \ E_1:\epsilon, \textit{cdr}:\textit{car}:\epsilon, E_1[\texttt{xs} = (1:2:3:4:\epsilon)])\\
\downarrow\\
(\textsc{call}\ 1:\textsc{call}\ 1:\textsc{env} \ E_1:\epsilon, (1:2:3:4:\epsilon):\textit{cdr}:\textit{car}:\epsilon, E_1[\texttt{xs} = (1:2:3:4:\epsilon)])\\
\downarrow\\
(\textsc{call}\ 1:\textsc{env} \ E_1:\epsilon, (2:3:4:\epsilon):car:\epsilon, E_1[\texttt{xs} = (1:2:3:4:\epsilon)])\\
\downarrow\\
(\textsc{env} \ E_1:\epsilon, 2:\epsilon, E_1[\texttt{xs} = (1:2:3:4:\epsilon)])\\
\downarrow\\
(\epsilon, 2:\epsilon, E_1)
\end{gather*}
\caption{\label{fig:lists} CSE execution for expression sequences and list processing.\ $E_1$\ \textrm{refers to the destructively updated environment}\ $E_0[\texttt{second} \leftarrow \textsc{clo} \ (\texttt{xs}) \ \texttt{(car (cdr xs))} \ E_1]$}
\end{figure}

The machine execution in Figure~\ref{fig:lists} demonstrates the handling of expression sequences and list processing using lists and the simple procedures \texttt{list}, \texttt{car} and \texttt{cdr} defined in $E_0$.

Compared to the SECD machine, the CSE machine uses a global operand stack (here called ``stash'') as opposed to starting each function application with a fresh operand stack. This allows us to combine the dump and the control of the SECD machine into a single control stack, with the help of the \textsc{env} instruction that restores the environment of the caller. Compared to the CEK machine, we avoid the need of continuations for explaining the basic execution structure and preserve the operand stack of the SECD machine, because it is needed anyway for succinctly explaining expression evaluation. The closure machine introduced by Kunze et al makes use of de Bruijn indices to represent environments as lists of closures, which leads them to combine the control and the environment into a single machine component. In contrast, the CSE machine makes parameters explicit and uses the traditional notion of environments from SICP and the SECD machine, which is more intuitive for novice computer scientists and does not require any preprocessing to generate indices.

The rules of the CSE machine in this and the next section are listed in the appendix for  reference.

\section{Extending the notional machine to cover \texttt{call/cc}}
\label{sec:call-cc}

Reified continuations are a useful feature in programming languages, providing a general and powerful means for programmers to manage program control flow and thereby enabling the creation of advanced control flow mechanisms such as exception handling. Notably, Scheme includes this as one of the key features of the full language.

The mental model for program execution provided by the CSE machine prepares the ground for explaining Scheme's continuations. Having defined
a program state as a triple consisting of the control, the stash and
the environment, it is but a small step to reify this program state into
a run-time value, akin to procedure closures.
\[
(C, \textsc{cont}\ C'\ S' \ E' : S, E)
\]
Here, the top item on the stash is a continuation that captures a machine state $(C', S', E')$ using the ternary constructor $\textsc{cont}$. Applying such a continuation will then establish the machine state that it captures.

We first require a mechanism to generate such continuations within the CSE machine. We provide a continuation-generating pseudoprocedure, $callcc$, bound in the initial environment to \texttt{call-with-current-continuation} and \texttt{call/cc}, as in Scheme. When $\textit{callcc}$ is applied, 
i.e. when the control has the instruction \textsc{call}\ 1 at the top and the stash has a procedure closure at the top followed by $callcc$, the machine generates a continuation that captures the current control (without the \textsc{call}\ 1 instruction), the current stash (without the procedure closure and $callcc$), and the current environment. The new control keeps \textsc{call}\ 1 at the top, but the new stash has the continuation at the top followed by the procedure closure. The effect is that the procedure argument of \texttt{call/cc} is called with the current continuation as argument.

\begin{gather*}
(\textsc{call}\ 1:C, \textsc{clo} \ p \ B \ E':\textit{callcc}:S, E) \rightarrow (\textsc{call}\ 1:C, \textsc{cont} \ C \ S \ E:\textsc{clo} \ p \ B \ E':S, E)  \\
\textrm{A call to}\ \mathit{callcc}
\end{gather*}


When applied using the $\textsc{call}$ instruction, a continuation will replace the current control, stash and environment with the values stored in the continuation, effectively returning the program execution back to the state at which the continuation was created.

\begin{gather*}
(\textsc{call}\ n:C, v_n:\ldots:v_1:\textsc{cont} \ C' \ S' \ E':S, E) \quad \rightarrow \quad (C', v_n:\ldots:v_1:S', E') \\
\textrm{Application of a continuation}\ \textsc{cont} \ C' \ S' \ E' \ \textrm{with arguments} \ v_1  \ldots  v_n. \\
\end{gather*}

\begin{figure}[t]
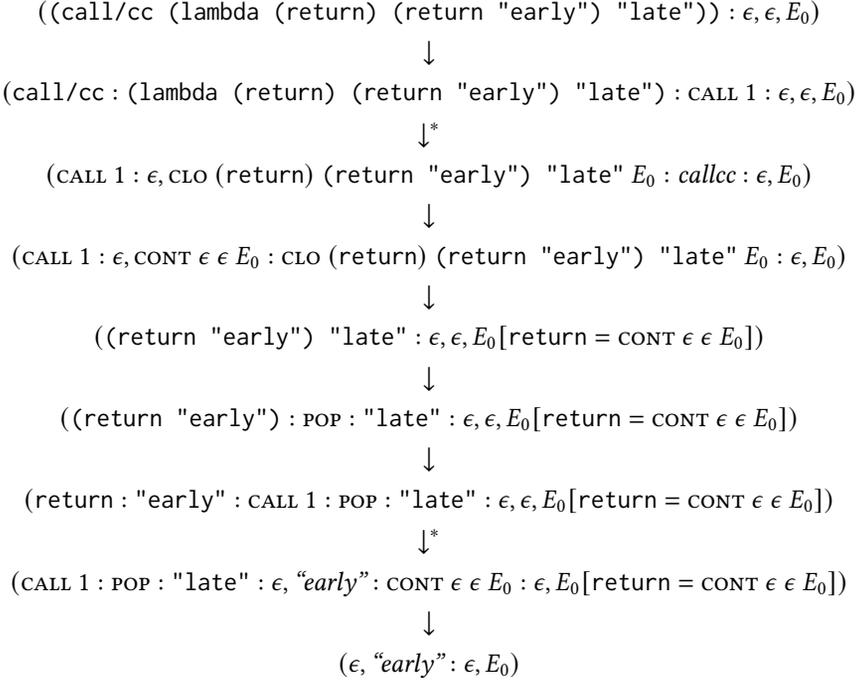

\begin{gather*}
(\texttt{(call/cc (lambda (return) (return "early") "late"))}:\epsilon, \epsilon, E_0)\\
\downarrow\\
(\texttt{call/cc}:\texttt{(lambda (return) (return "early") "late")}:\textsc{call}\ 1:\epsilon, \epsilon, E_0)\\
\downarrow^*\\
(\textsc{call} \ 1:\epsilon, \textsc{clo} \ (\texttt{return}) \ \texttt{(return "early") "late"} \ E_0:\textit{callcc}:\epsilon, E_0)\\
\downarrow\\
(\textsc{call} \ 1:\epsilon, \textsc{cont} \ \epsilon \ \epsilon \ E_0:\textsc{clo} \ (\texttt{return}) \ \texttt{(return "early") "late"} \ E_0:\epsilon, E_0)\\
\downarrow\\
(\texttt{(return "early") "late"}:\epsilon, \epsilon, E_0[\texttt{return} = \textsc{cont} \ \epsilon \ \epsilon \ E_0])\\
\downarrow\\
(\texttt{(return "early")}:\textsc{pop}:\texttt{"late"}:\epsilon, \epsilon, E_0[\texttt{return} = \textsc{cont} \ \epsilon \ \epsilon \ E_0])\\
\downarrow\\
(\texttt{return}:\texttt{"early"}:\textsc{call} \ 1:\textsc{pop}:\texttt{"late"}:\epsilon, \epsilon, E_0[\texttt{return} = \textsc{cont} \ \epsilon \ \epsilon \ E_0])\\
\downarrow^*\\
(\textsc{call} \ 1:\textsc{pop}:\texttt{"late"}:\epsilon, \textit{``early''}:\textsc{cont} \ \epsilon \ \epsilon \ E_0:\epsilon, E_0[\texttt{return} = \textsc{cont} \ \epsilon \ \epsilon \ E_0])\\
\downarrow\\
(\epsilon, \textit{``early''}:\epsilon, E_0)
\end{gather*}
\caption{\label{fig:return} Implementing explicit return using continuation}
\end{figure}

Figure~\ref{fig:return} demonstrates how a continuation is applied. The example employs a continuation as a means to return early from procedure execution, a common feature of modern languages.

\section{Implementation}
\label{sec:implementation}

Our implementation of the CSE machine extends the open-source system Source Academy~\cite{sa} that lets the readers of SICP JS run programs in the browser. The Source Academy organization also provides a SICP comparison edition~\cite{sicp-compare} that offers a side-by-side view of the two editions. We intend to contribute an interactive version of the original SICP, and our implementation of the CSE machine was developed to run and visualize the Scheme programs of SICP in the browser. 

Scheme is implemented in Source Academy via translation to JavaScript. This approach was chosen as it allows the Scheme implementation to utilise all Source Academy features that were originally designed for SICP~JS, including a stepper tool~\cite{splash-e-stepper-2021} for SICP's substitution model and modules for the picture language of SICP 2.2.4. Source Academy implements proper tail calls with a JavaScript-to-JavaScript transpiler that uses trampolines~\cite{trampoline}.

In the current release of Source Academy, Scheme and JavaScript programs can be run and visualized with a common CSE machine implementation. The Scheme implementation uses a smaller instruction set that closely follows the idealized CSE machine described in this paper. However, transpilation to JavaScript necessitated deviations, which are discussed below.

\begin{figure}[t]
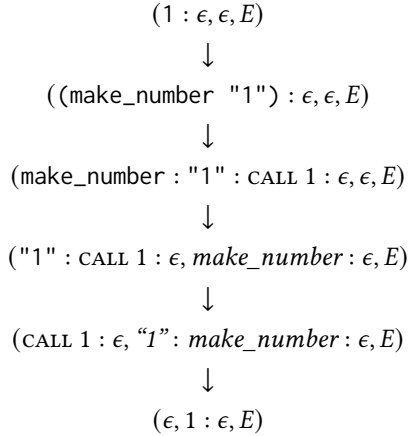

\begin{gather*}
(\texttt{1}:\epsilon, \epsilon, E) \\
\downarrow \\
(\texttt{(make\_number "1")}:\epsilon, \epsilon, E) \\
\downarrow \\
(\texttt{make\_number}:\texttt{"1"}:\textsc{call}\ 1:\epsilon, \epsilon, E) \\
\downarrow \\
(\texttt{"1"}:\textsc{call}\ 1:\epsilon, \textit{make\_number}:\epsilon, E) \\
\downarrow \\
(\textsc{call}\ 1:\epsilon, \textit{``1''}:\textit{make\_number}:\epsilon, E) \\
\downarrow \\
(\epsilon, 1:\epsilon, E)
\end{gather*}
\caption{\label{numerical} Derivation for the transition $(\texttt{1}:\epsilon, \epsilon, E) \rightarrow^* (\epsilon,1:\epsilon, E)$}
\end{figure}

While our Scheme implementation internally implements the full numerical tower by using a special data representation for numbers, our CSE machine employs simple procedures to produce this representation on the stash. For example, Figure~\ref{numerical} shows how the CSE machine employs the simple procedure $\textit{make\_number}$ to produce a numerical value on the stash that corresponds to the trivial Scheme program \texttt{1}.
Similarly, Scheme symbols are created indirectly by using the \texttt{string->symbol} procedure. Deviations from expected behaviour through invalid input to these procedures are prevented by the Scheme parser, which invalidates programs with ill-formed numbers.

One key difference between Scheme and JavaScript is the \texttt{return} syntax in JavaScript that allows early exits from functions. To remove unexecuted control items after an early exit using a return statement, the \textsc{return} and corresponding \textsc{mark} instructions designate a range of control items to delete. In Scheme, procedures always terminate with the final statement, and so transpiled Scheme programs are set with a \texttt{return} statement at the end of each procedure body. Hence, the effect of these instructions, while visible in the common CSE machine, may be ignored as there are never any instructions to be removed.

The CSE machine is implemented as a generator function that outputs program states until there are no further items on the control stack to execute or until the number of generated states exceeds a set state limit provided as a user-specified setting in the CSE machine. The program execution is initiated by the event handler of the ``Run'' button in the web page. Each machine state is assigned a state number, which allows users to request stepping through programs forwards and backwards. Backward stepping is currently implemented by recomputing the program state from the beginning up to the specified state number. This means that the machine produces a coherent behavior only when the given program is deterministic. The state limit also handles infinite loops, preventing non-responsiveness or hanging in the browser platform, as the program terminates execution after a specified step limit. However, this limit can be changed if users expect their programs to execute with more steps.

The CSE machine is being continuously developed. A possible improvement is migrating program execution from the browser tab's event handler to web workers for program execution, which would allow the browser to remain responsive even if the CSE machine is still executing a program.

\section{Visualization}
\label{sec:visualization}

\begin{figure}[btp]
    \centering
    \includegraphics[width=1.0\linewidth]{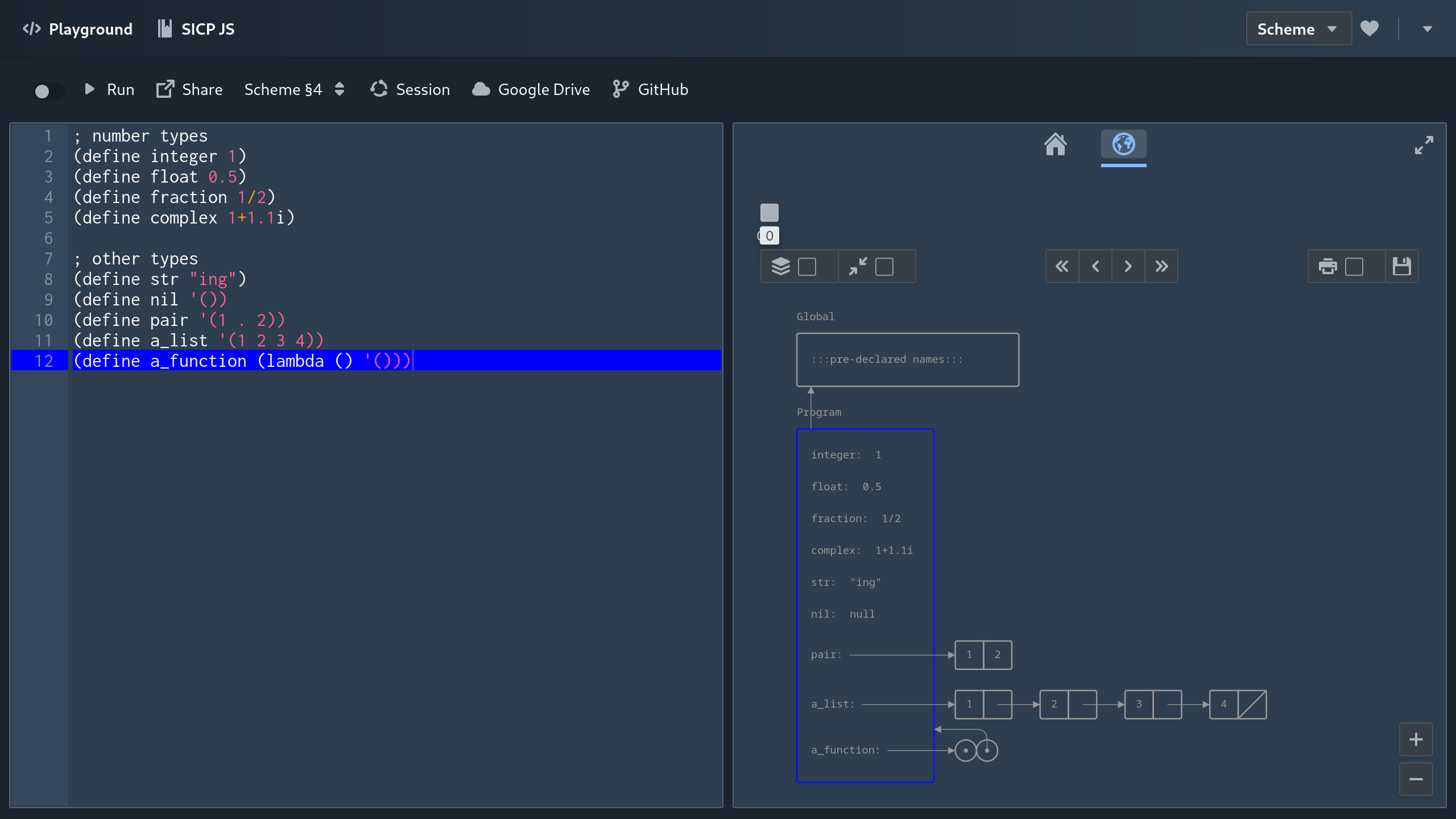}
    \caption{\label{fig:env-lists}Scheme data types and environments shown in the environmental visualizer}
\end{figure}

The textual representation of the CSE machine transitions presented in Section~\ref{sec:notional} is not evocative and intuitive enough to be used as an effective teaching tool. We need an appealing interactive visualization that can serve as a reference to students. Our visualization of the CSE machine builds on an open-source environment visualizer~\cite{env-visu-2023}, which is a tool in Source Academy that allows learners to view run-time data structures used in the execution of a JavaScript program, following the principles of the environment model introduced in SICP. Due to the transpiled nature of Scheme programs in our implementation, we are able to represent Scheme programs using the same environment visualizer. Additionally, we visualize changes in the environment as the program executes by providing a slider, giving learners the illusion of stepping forwards and backwards through program execution as discussed in the previous section.  Figure~\ref{fig:env-lists} shows how the Scheme adaptation of the Source Academy environment visualizer represents SICP's box-and-pointer diagrams, closures, and environments.

\begin{figure}[tbh]
    \centering
    \includegraphics[width=1.0\linewidth]{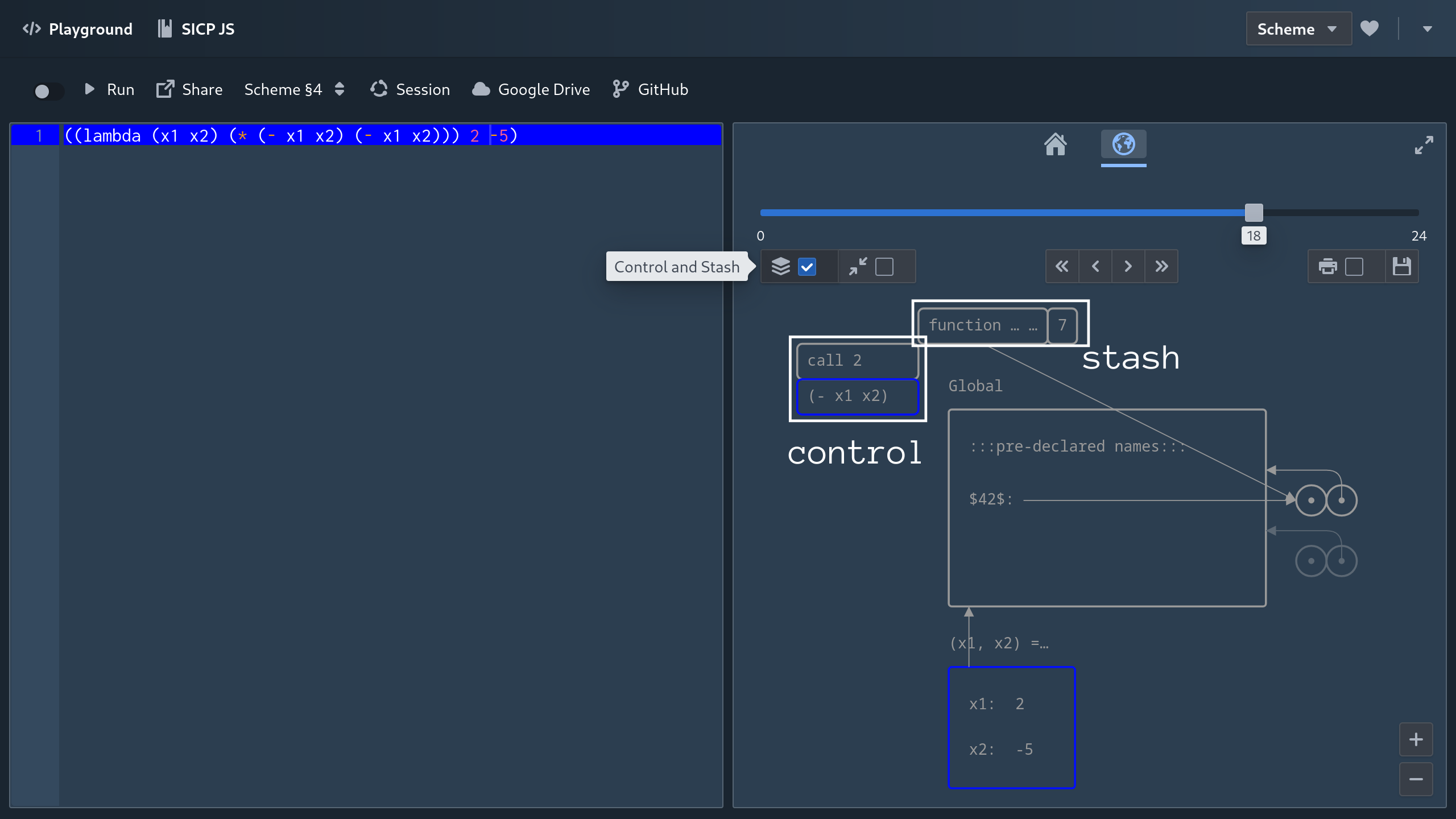}
    \caption{\label{fig:enter-label}The control and stash components of the CSE machine during program execution}
\end{figure}

We built the CSE machine visualizer as an extension of the environment visualizer~\cite{env-visu-2023}, allowing learners to view the explicit step by step execution of their programs in addition to viewing environment diagrams as well as procedure objects for the program that are in line with SICP representations. Learners can choose to visualize the control and stash components of the CSE machine by toggling a button as shown in Figure~\ref{fig:enter-label}. The control and stash components appear on the upper left of the environment model, showing control items such as Scheme code or instructions on the control, as well as data types such as primitive data types, or procedure closures, represented as procedure objects on the stash.

\begin{figure}[tbp]
    \centering
    \includegraphics[width=1.0\linewidth]{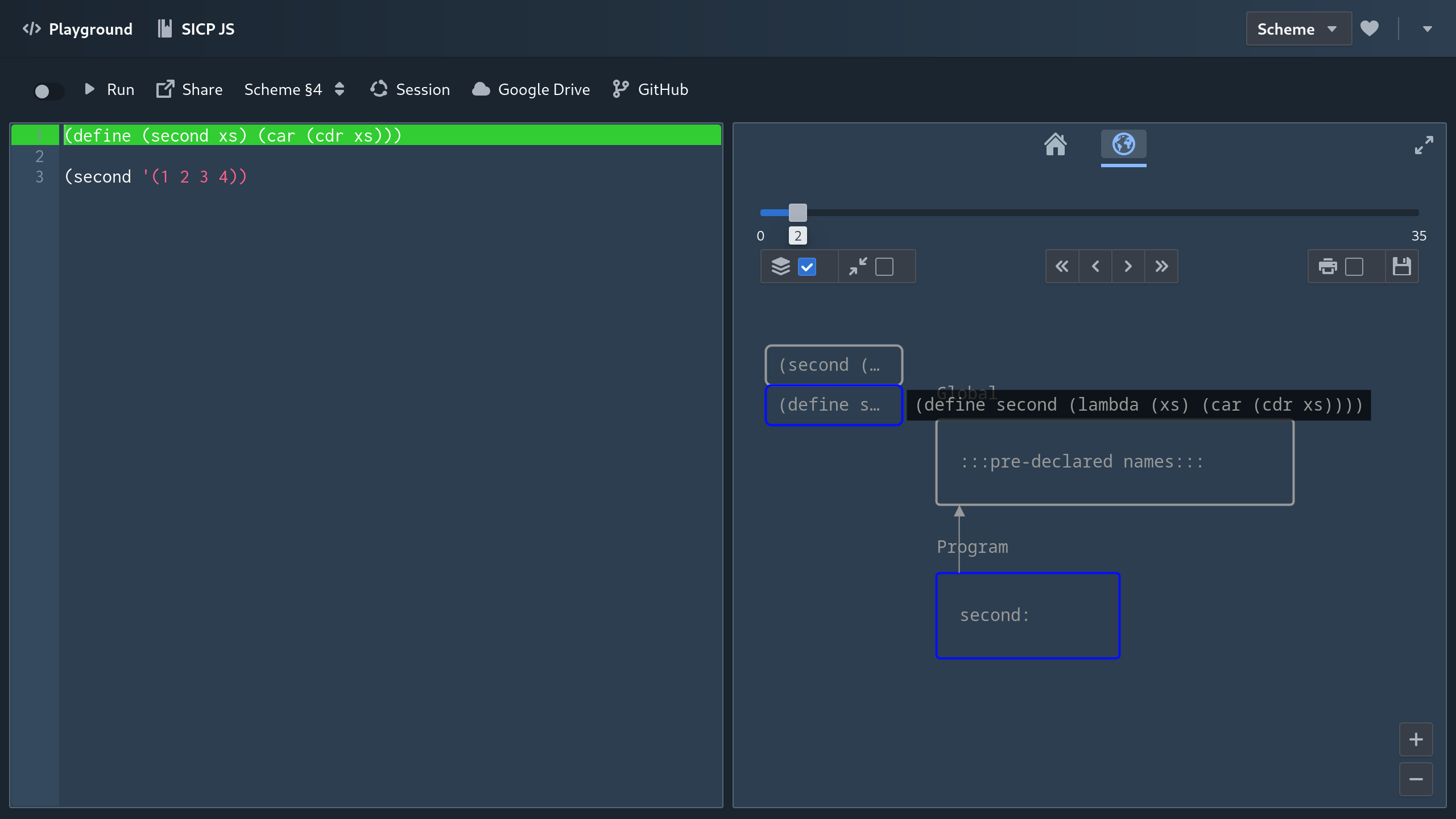}
    \caption{\label{fig:highlight}A highlighted region of the program associated with the selected control component}
   
\end{figure}

To aid in clarity, procedure closures on the stash point to the proper procedure object that they represent, while control components also highlight the region of the program that they are associated with, as shown in Figure~\ref{fig:highlight}. At any program state, the exact line of the program that is being executed is highlighted blue.

\begin{figure}[tp]
    \centering
    \includegraphics[width=0.95\linewidth]{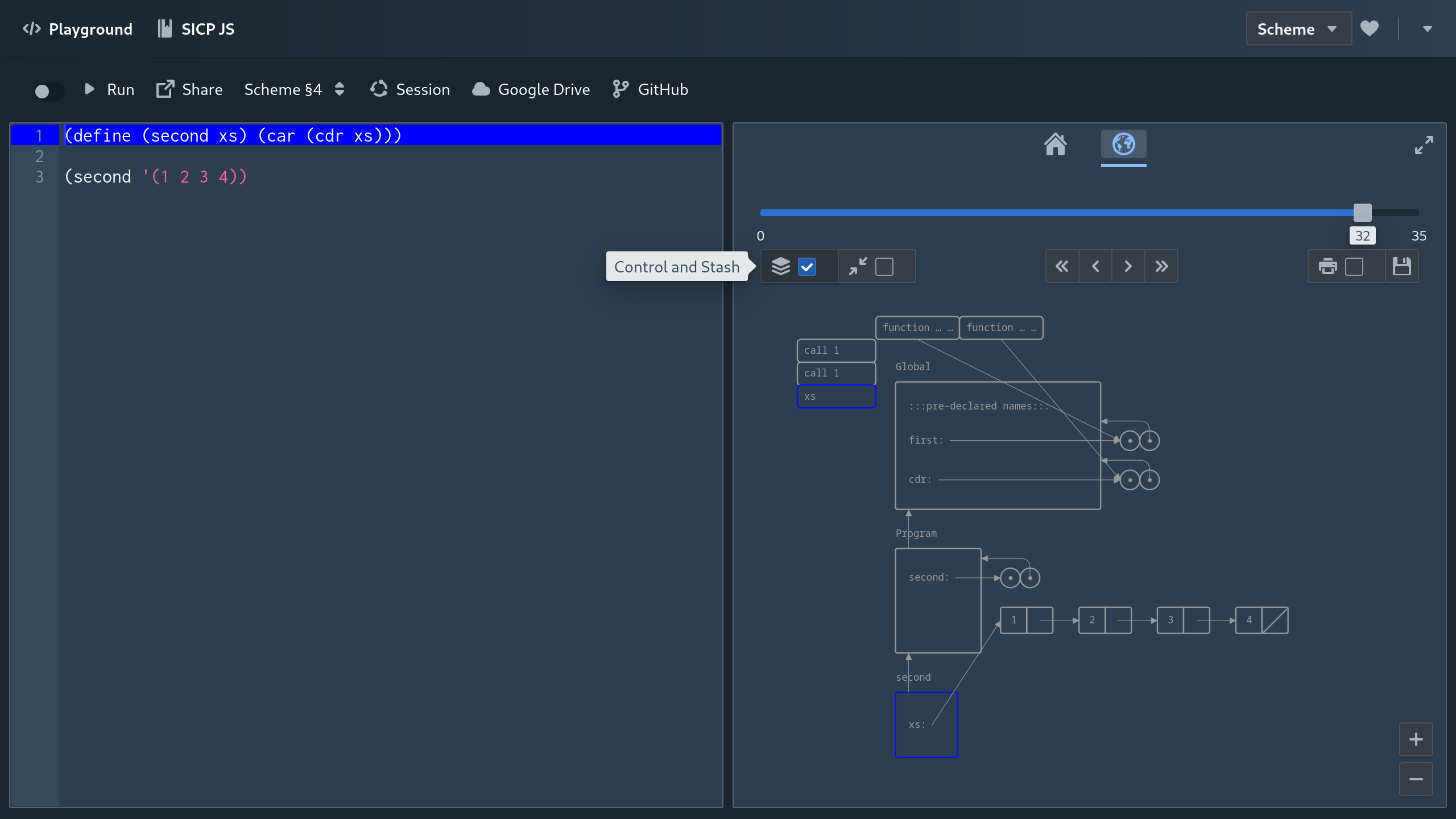}
    \caption{\label{fig:viz-lists} A program operating on lists, shown during execution}
\end{figure}

In Figure~\ref{fig:viz-lists}, we use the final example of Section~\ref{sec:notional} to demonstrate the effectiveness of the visualisation in demonstrating the mechanics of the CSE machine. We are able to view data types such as procedure closures or cons cells, relevant built-in functions, namely \texttt{cdr} and \texttt{first} / \texttt{car}, view the current environment, which is coloured blue.

\begin{figure}[tpb]
    \centering
    \includegraphics[width=0.87\linewidth]{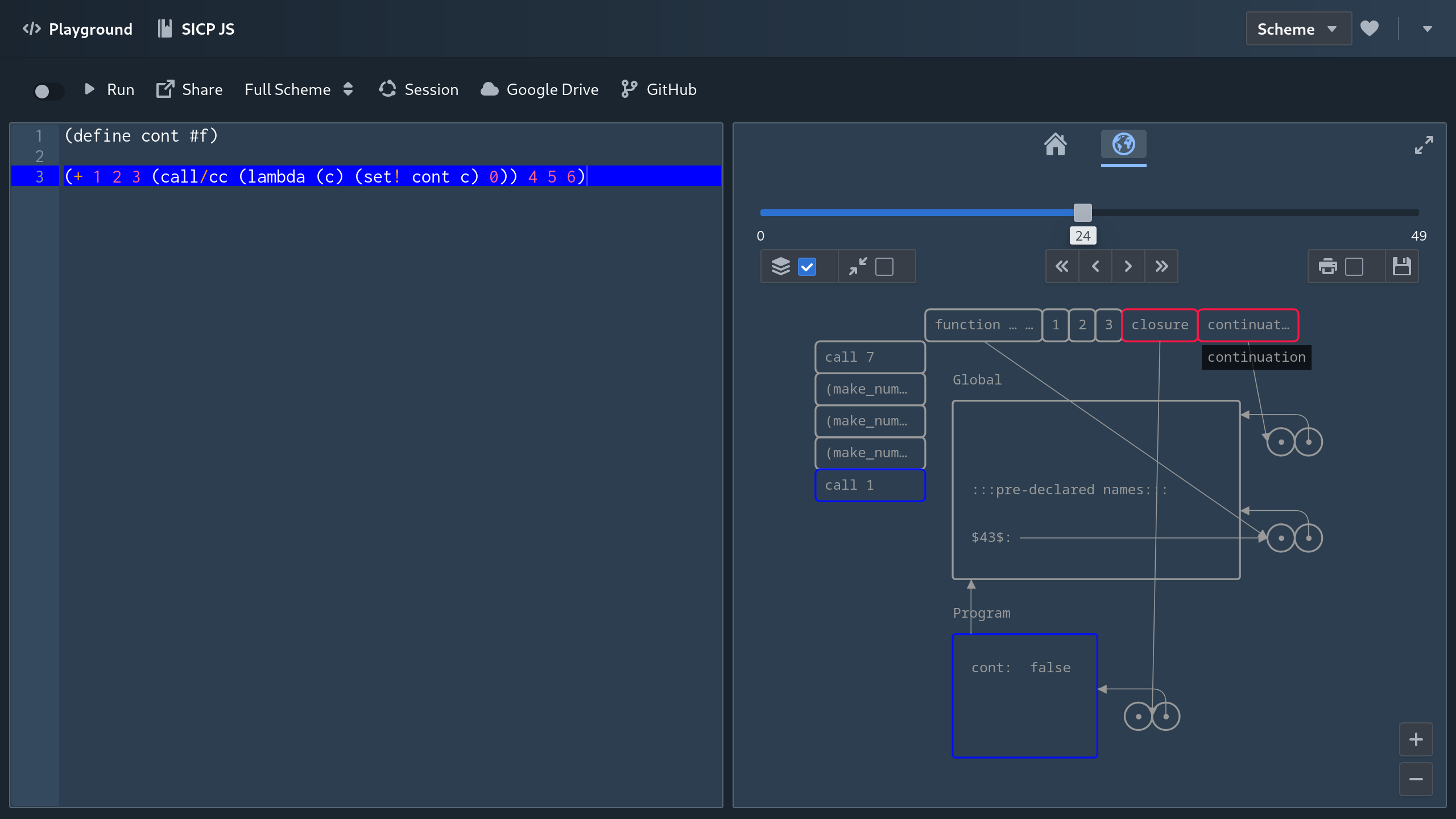}
    \caption{\label{fig:viz-cont}A continuation represented within the CSE machine. At the time of writing, the environment component is non-functional and displays an incorrect environment as a proper representation for continuations has not been implemented yet}
\end{figure}

At the time of writing, continuations are visible within the CSE machine, and can be identified separately in the stash, but there is no special diagrammatic representation for the control and stash within a continuation; see Figure~\ref{fig:viz-cont}.

\begin{figure}[t]
    \centering
    \includegraphics[width=0.9\linewidth]{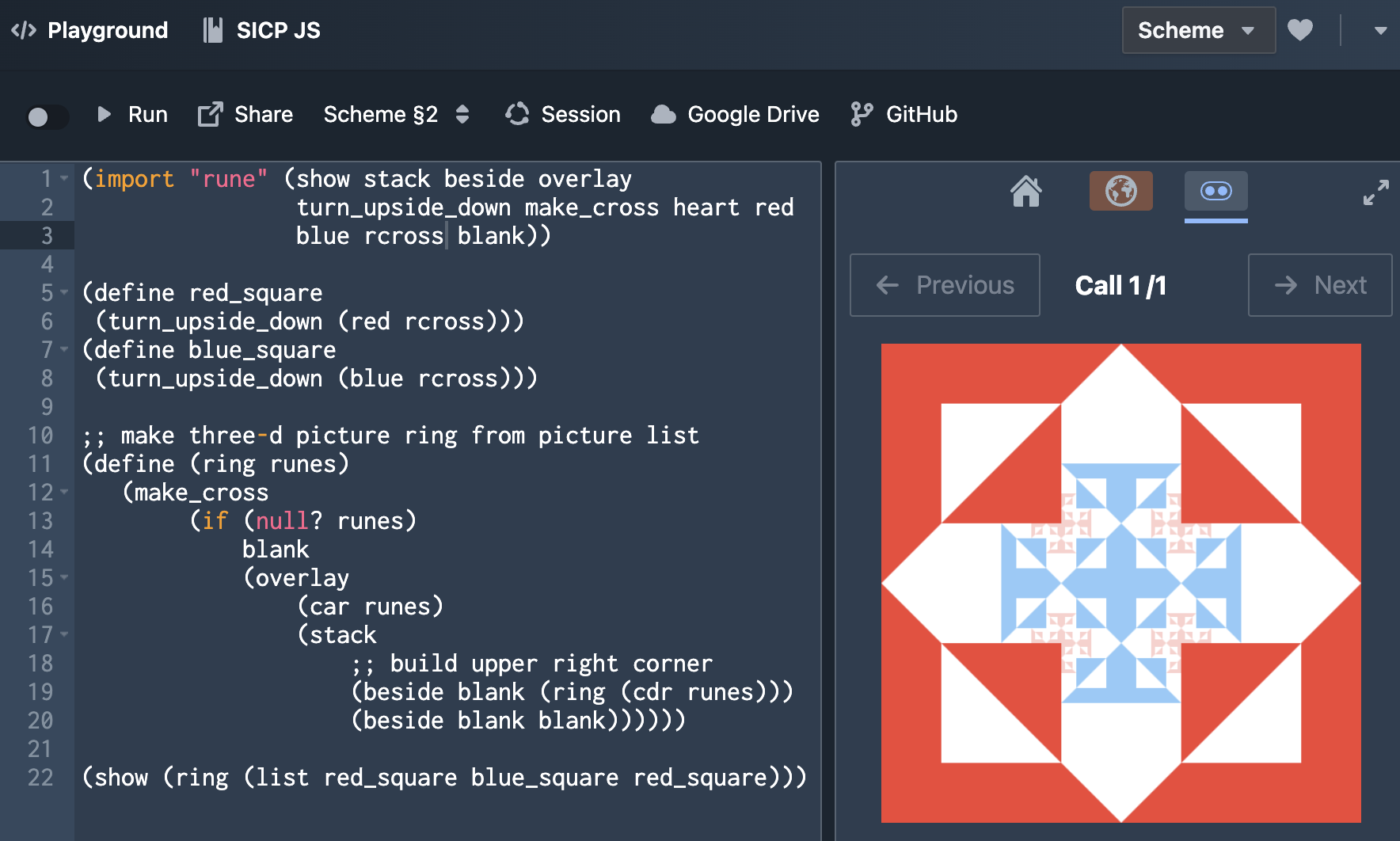}
    \caption{\label{fig:runes}Supporting the picture language of SICP 2.2.4 in Scheme in the browser}
\end{figure}

A central motivation for Source Academy to be browser-based was to provide educators with access to the media-rich environment of the web. Today, Source Academy comes with a wide range of \emph{modules}~\cite{modules} that provide support for SICP's picture language, sound~\cite{splash-e-teachable-2021} and video processing, robotics~\cite{splash-e-ruggedizing-2021} and many other experiential learning components. As an example, based on the support of the picture language of SICP Section 2.2.4 in Source Academy, it was trivial to provide a Scheme implementation, as illustrated in Figure~\ref{fig:runes}.

\section{Conclusion
}
\label{sec:conclusion}

In Section~\ref{sec:notional}, we have shown that the environment model of SICP can be extended to become a fully-fledged notional machine that we call the \emph{CSE machine}, a simplified and modified version of the classic SECD machine. The implementation and visualization of the CSE machine presented in Sections~\ref{sec:implementation} and~\ref{sec:visualization} make use of the web-based infrastructure of Source Academy, a tool built~\cite{anderson_sigcse2023} for teaching CS1101S, an introductory computer science course at the National University of Singapore. CS1101S follows SICP JS, a JavaScript adaptation of SICP. The Fall 2023 edition of CS1101S used the JavaScript version of the CSE machine and its visualization in the Source Academy. Student and tutor feedback indicates that the CSE machine is a useful extension of SICP's environment model. In the future, CS1101S will give more prominence to the coherent mental model for program execution provided by the CSE machine. 

Modern languages such as Python and JavaScript inherit some of their computational concepts from Scheme. Thus, the transpiler approach described in Section~\ref{sec:implementation} opens the opportunity of implementing these languages along with their ancestor Scheme in a common programming environment.
Based on our work so far, it appears feasible that the pedagogical content of SICP can serve educators who use any of these languages, not just Scheme and JavaScript. Our vision is to extend Source Academy to become a platform where educators can choose their favorite language among Scheme, JavaScript, Python, Lua, and others, and where they find the corresponding version of SICP, along with tools such as the CSE machine and experiential learning components that all work consistently across all languages, without students needing to install any programming language tools and without the need to redevelop these tools for the different languages.

Section~\ref{sec:call-cc} showed an intriguing avenue for bridging the pedagogical gap between basic sequential languages commonly covered in first-year courses and more advanced control structures such as explicit return and try/catch-style exception handling. Scheme's \texttt{call/cc} construct subsumes these and many other control structures. We showed that \texttt{call/cc} can be explained within the framework of the CSE machine. That means that instructors can choose to introduce these control structures in their courses and use the CSE machine as a coherent mental model.



\bibliographystyle{ACM-Reference-Format}
\balance
\bibliography{shared}


\appendix

\section*{Appendix: Rules of the CSE Machine}
\label{sec:appendix}

\begin{table}[hbt!]
\centering
    \begin{tabular}{|p{30mm}|p{49mm}|p{49mm}|}
        \hline
        \multicolumn{3}{|c|}{Expression Decomposition Rules} \\
        \hline
        Rule & Initial State & Final State\\
        \hline
        \mbox{Decompose \textit{n}-ary}  procedure call  & $(\texttt{(} V_0\ V_1\ldots  V_n \texttt{)}:C, S, E)$ & $(V_0: V_1:\ldots:V_n:\textsc{call}\ n:C, S, E)$ \\
        \hline
        Construct closure  & $(\texttt{(lambda\ (}x_1 \ldots  x_n\texttt{)} \ B \texttt{)}:C, S, E)$ & $(C, \textsc{clo} \ (x_1,\ldots, x_n) \ B \ E:S, E)$ \\
        \hline
        Decompose variable declaration  & $(\texttt{(define} \ x \ V \texttt{)}:C, S, E)$ & $(V:\textsc{asgn} \ x:C, S, E)$ \\
        \hline
        Decompose variable assignment  & $(\texttt{(set!} \ x \ V \texttt{)}:C, S, E)$ & $(V:\textsc{asgn} \ x:C, S, E)$ \\
        \hline
        Decompose conditional expression  & $(\texttt{(if} \ V \ A \ B \texttt{)}:C, S, E)$ & $(V:\textsc{branch}\ A \ B:C, S, E)$ \\
        \hline
        Decompose expression sequence  & $(V_1  \ldots  V_n:C, S, E)$ & $(V_1:\textsc{pop}:\ldots:V_n:C, S, E)$ \\
        \hline
        Decompose \texttt{begin} expression  & $(\texttt{(begin} \ V_1  \ldots  V_n \texttt{)}:C, S, E)$ & $(V_1:\textsc{pop}:\ldots:V_n:C, S, E)$ \\
        \hline
    \end{tabular}
\end{table}

\begin{table}[hbt!]
\centering
    \begin{tabular}{|p{30mm}|p{49mm}|p{49mm}|}
        \hline
        \multicolumn{3}{|c|}{Reduction Rules} \\
        \hline
        Rule & Initial State & Final State\\
        \hline
        Evaluate primitive & $(v:C, S, E)$ & $(C, v:S, E)$ \\
        \hline
        Lookup variable & $(x:C, S, E)$ & $(C, E.x:S, E)$ \\
        \hline
        \mbox{Apply operator or} simple procedure & $(\textsc{call}\ n:C, v_n:\ldots:v_1:f:S, E)$ & $(C, f(v_1,\ldots,v_n):S, E)$ \\
        \hline
        Apply closure  & \begin{minipage}{50mm}
        $(\textsc{call}\; n:C,$\newline 
    $v_n:\ldots:v_1:\textsc{clo} (x_1\ldots x_n)\ B\ E':S,$\newline
        $ E)$
                \end{minipage}
        & 
        \begin{minipage}{50mm}
        $(B:\textsc{env}\ E:C,$\newline
        $S,$\newline
        $E'[x_1 = v_1]\ldots[x_n = v_n])$
        \end{minipage}
        \\
        \hline
        Apply \textit{callcc} & 
        \begin{minipage}{50mm}
        $(\textsc{call}\ 1:C,$\newline
        $\textsc{clo} \ p \ B \ E':\textit{callcc}:S, E)$
        \end{minipage}
        & 
        \begin{minipage}{50mm}
        $(\textsc{call}\ 1:C,$\newline
        $\textsc{cont} \ C \ S \ E:\textsc{clo} \ p \ B \ E':S, E)$
        \end{minipage}
        \\
        \hline
        Apply continuation 
        & 
        \begin{minipage}{50mm}
        $(\textsc{call}\ n:C,$ \newline
        $v_n:\ldots:v_1:\textsc{cont} \ C' \ S' \ E':S, E)$
        \end{minipage}& $(C', v_n:\ldots:v_1:S', E')$\\
        \hline
        Restore environment & $(\textsc{env}\ E':C, S, E)$ & $(C, S, E')$ \\
        \hline
        \mbox{Assign variable to} value & $(\textsc{asgn} \ x:C, v:S, E)$ & $(C, v:S, E[x \leftarrow v])$ \\
        \hline
        Branch$\;$to$\;$consequent (truthy $v$)  & $(\textsc{branch}\ A \ B:C, v:S, E)$ & $(A:C, S, E)$ \\
        \hline
        Branch to alternative (falsy $v$) & $(\textsc{branch}\ A \ B:C, v:S, E)$ & $(B:C, S, E)$ \\
        \hline
        \mbox{Remove unused} value & $(\textsc{pop}:C, v:S, E)$ & $(C, S, E)$ \\
        \hline
    \end{tabular}
\end{table}

\end{document}